\documentclass[a4paper,11pt]{article}
\usepackage{jcappub}
\usepackage{newtxtext,newtxmath}

\usepackage{lineno}
\usepackage{ae,aecompl}
\usepackage{graphicx}	
\usepackage{amsmath}	
\usepackage{threeparttable}
\usepackage{subcaption}
\usepackage{caption}
\usepackage{color}
\usepackage[dvipsnames]{xcolor}
\usepackage[normalem]{ulem}
\usepackage{booktabs}
\usepackage{multirow}

\newcommand{\HI}{H\,{\sc i}}

\def \link_col{blue}

\newcommand{\gray}{$\gamma$-ray }

\title{\boldmath Cosmic ray density variations in nearby giant molecular clouds}

\author[a,b,c]{Jiahao Liu,}
\author[a,b,c]{Bing Liu}
\author[a,b,c]{and Ruizhi Yang}
\affiliation[a]{Deep Space Exploration Laboratory/School of Physical Sciences, University of Science and Technology of China, Hefei 230026, China}
\affiliation[b]{CAS Key Laboratory for Research in Galaxies and Cosmology, Department of Astronomy, School of Physical Sciences, University of Science and Technology of China, 
Hefei, Anhui 230026, China}
\affiliation[c]{School of Astronomy and Space Science, University of Science and Technology of China, 
Hefei, Anhui 230026, China}

\emailAdd{lbing@ustc.edu.cn}

\abstract{In this paper, we analyzed  12 years of {\sl Fermi} LAT gamma-ray data towards three nearby giant molecular clouds (GMCs),
i.e., R~CrA, Chamaeleon, and Lupus. We calibrated the gas column density of these regions by using the Planck dust opacity map as well as the Gaia extinction map. With both the gamma-ray observations and gas column density maps, we derived the cosmic ray densities in the three GMCs. We found the derived CR spectra have almost the same shape but significantly different normalizations, which may reflect that the distributions of CRs in the vicinity of solar systems are inhomogeneous.}


\begin{document}
\maketitle
\flushbottom

\section{Introduction}

Cosmic rays (CRs), as one of the most important ingredients in the interstellar medium (ISM),  dominate the heating and ionization of gas inside molecular clouds, regulate the star-forming rate, and play a leading role in the astrochemistry process.  Composed of mainly charged particles, CRs are well mixed in the Galactic magnetic field and lost the information of their acceleration sites. Furthermore, it is debatable whether or not the CR density measured in the solar system is a good representative of that in the whole Galaxy. In this regard, gamma rays, the secondary products of interactions between CRs and the ISM, provide us with crucial information about the distribution of CRs in the Galaxy. Giant molecular clouds (GMCs), as the CR barometers \cite{aharonian01}, are considered  the ideal sites to study CRs. The gamma rays in GMCs have already been extensively studied after the Launch of {\sl Fermi} satellite, and similar CR densities and spectra inside the GMCs in the Gould Belt are derived \citep{neronov12,yang14}.

In this paper, we analyzed 12 years of {\sl Fermi} Large Area Telescope (LAT) observations towards three GMCs, including R CrA, Chamaeleon, and Lupus. Compared with previous studies\citep{yang14,neronov12,peron21}, in addition to the updated {\sl Fermi}-LAT analysis with more exposure, we also applied the method described above for a new estimation of the gas mass, which may lead us to new results.
 Indeed, in GMCs the gamma rays are mainly produced by the interaction of CRs with the ambient gas. Thus, the gamma-ray spectra can be used to determine the spectral shape of the parent CRs, meanwhile, the absolute normalization of CR density depends also on the gas mass. In the former work, both CO and dust opacity maps are used to trace the gas \citep{yang14,neronov12,peron21}. However, the CO emission may underestimate the total gas due to the existence of the well-known CO dark gas \citep{grenier05}.  The dust opacity map is free of the "dark gas" problem, but the ratio between dust opacity and gas column may vary significantly in different positions of the sky \citep{2013Planck}. In this regard, the extinction map may be a more reliable tracer of gas, since the dust absorption depends less on the environment compared with the dust emission, which was used to derive the dust opacity. However, in very dense regions such as the core of GMCs, extinction may not be measured. Thus in this work, we derived the linear relation between the Gaia extinction map \citep{2022Delchambre} and the Planck dust opacity in different GMCs. After that we derived the ratio between Planck dust opacity and gas column density, assuming the ratio between extinction and gas column density is constant. In this way, we derived  more reliable gas column density maps, which were used to study the CR density in different clouds.

\section{{\sl Fermi}-LAT  Data Analysis}
\label{sec:gamma}
In order to study the gamma-ray emission of the GMCs, we collected 12 years of {\sl Fermi}-LAT Pass 8 data, which is the latest data release. We used the Fermitools from Conda distribution
and applied the latest version of the instrument response functions ( $P8R3\_SOURCE\_V3$). We chose $15^{\circ}\times15^{\circ}$ regions centered at the positions of R CrA (R.A.=285.89$^{\circ}$, Dec.=-36.50$^{\circ}$), Chamaeleon (R.A.=236.19$^{\circ}$, Dec.=-34.15$^{\circ}$), and Lupus (R.A.=178.75$^{\circ}$, Dec.=-78.00$^{\circ}$) 
as the regions of interest (ROI). For Lupus and R~CrA, we performed the analysis under the equatorial coordinate system, but for Chamaeleon, we chose the galactic coordinate due to the aberration caused by the high declination.

\begin{figure*}[htbp]
    \centering
    \begin{subfigure}[t]{0.3\linewidth}
        \centering
        \includegraphics[width=\linewidth]{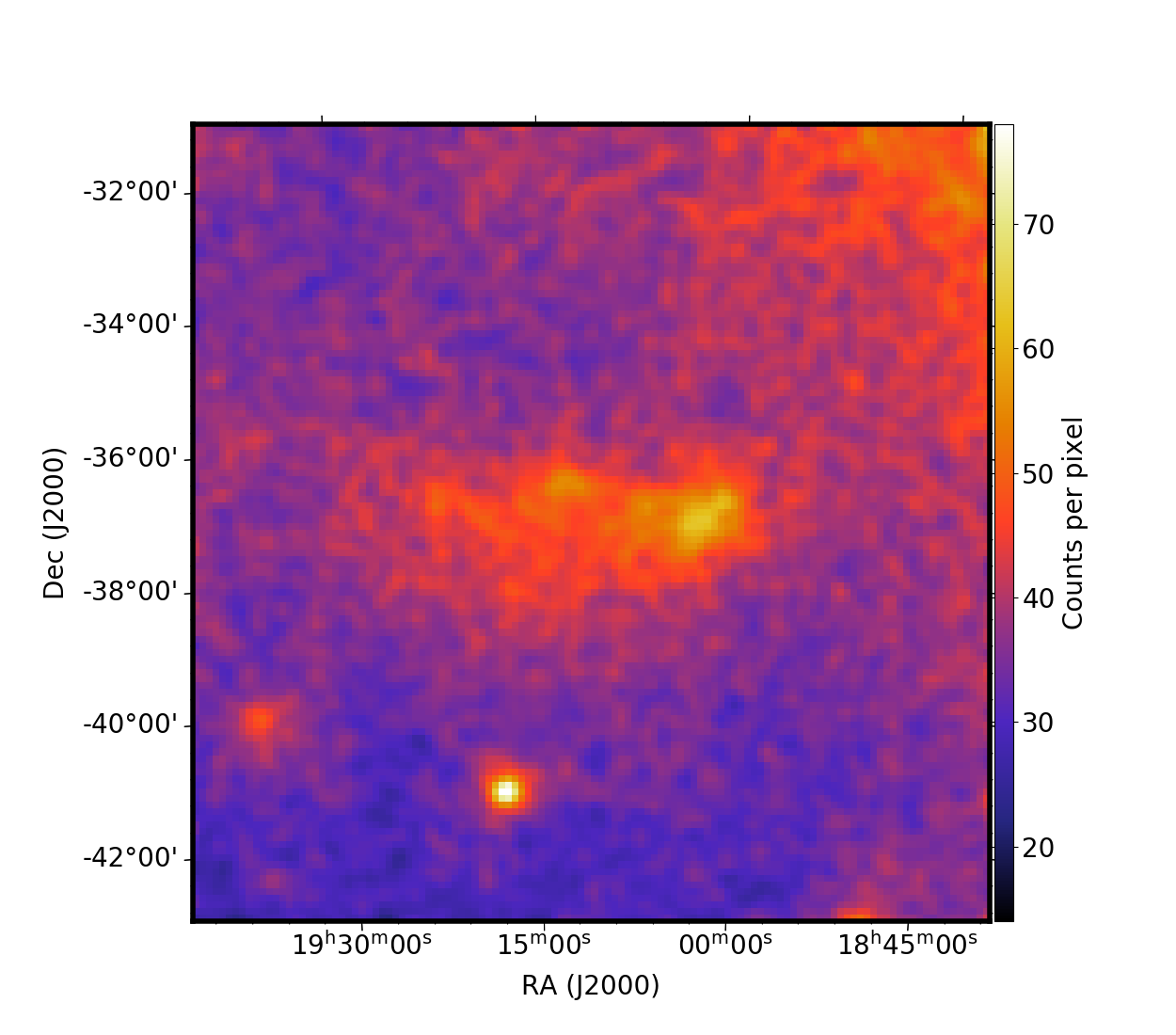}
        \caption{R~CrA }\label{counts:1}
    \end{subfigure}
    \quad
    \begin{subfigure}[t]{0.3\linewidth}
        \centering
        \includegraphics[width=\linewidth]{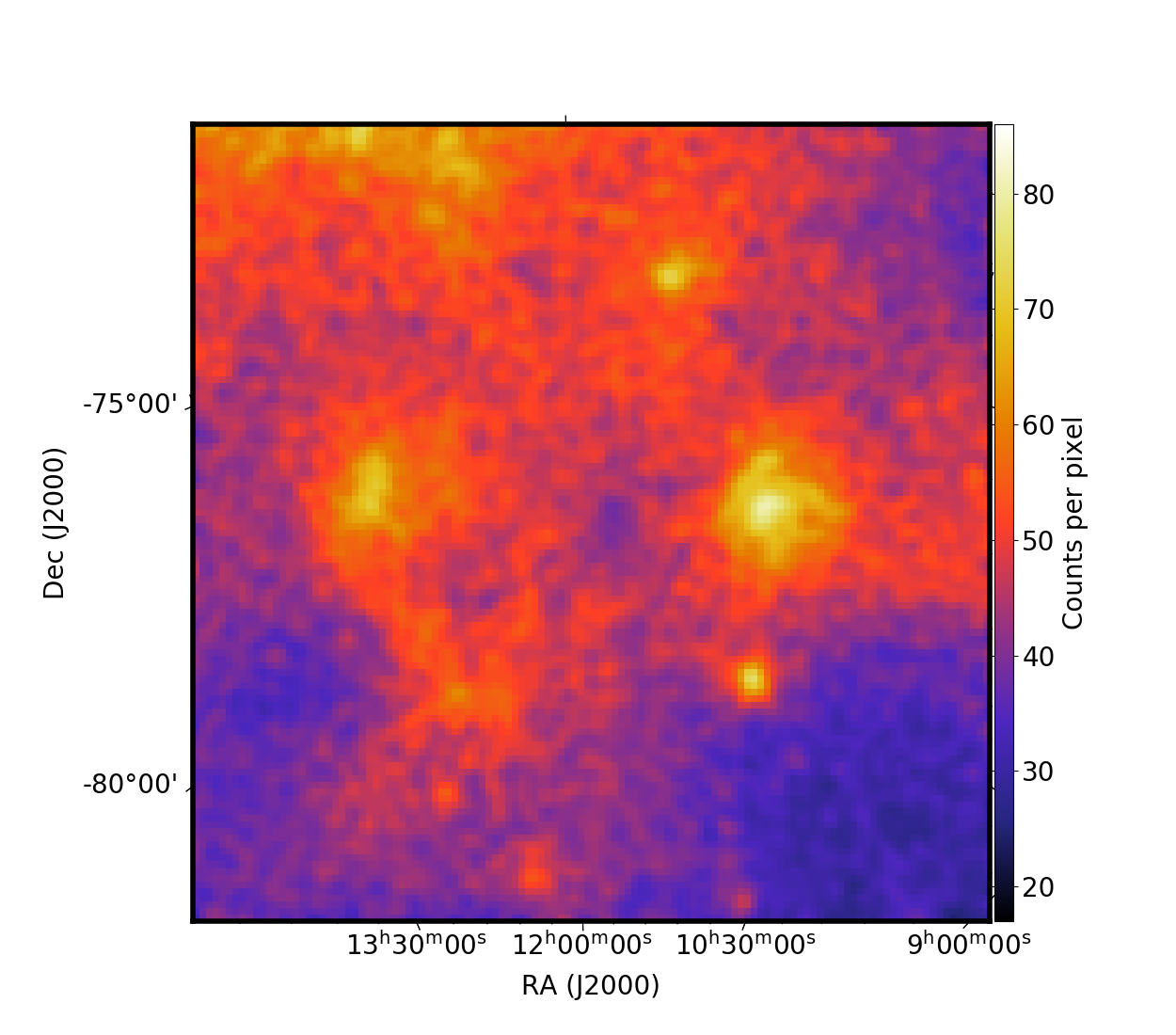}
        \caption{Chamaeleon}\label{counts:2}
    \end{subfigure}
    \begin{subfigure}[t]{0.3\linewidth}
        \centering
        \includegraphics[width=\linewidth]{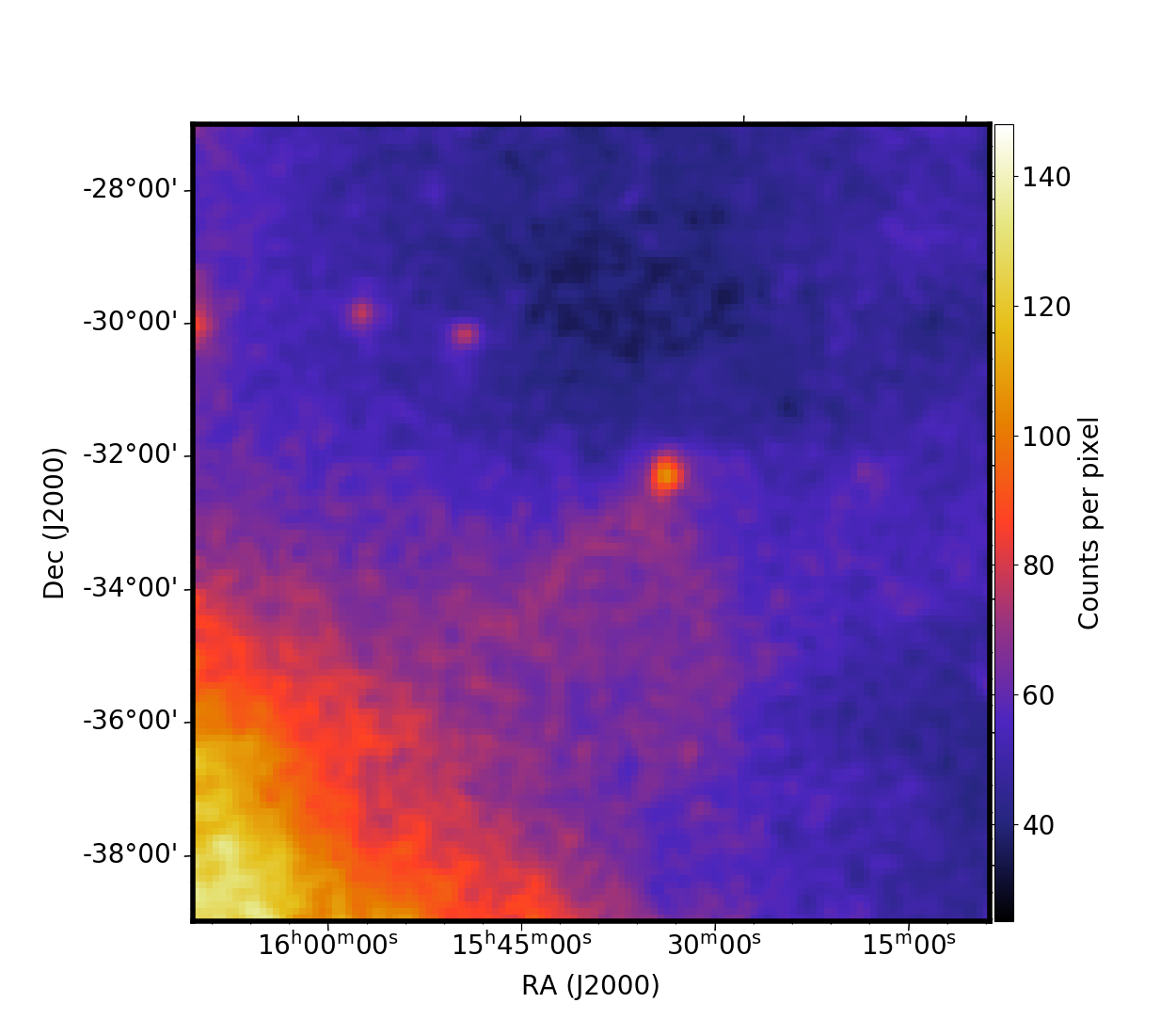}
        \caption{Lupus}\label{counts:3}
    \end{subfigure}
    \caption{Observed gamma-ray counts maps of the GMCs, smoothed with Gaussian filter of 0.3$^{\circ}$. }\label{fig:counts} 
\end{figure*}

We used {\sl gtselect}  to select the "source" class events within the energy range of 0.1-500~GeV to conduct our research, and applied the data filtrating equation $(\text{DATA}\_\text{QUAL}>0)\&\&(\text{LAT}\_\text{CONFIG}==1)$ to choose the good time intervals. Additionally, we only included the events whose zenith angles are less than 90$^{\circ}$ to filter out the background contamination from the Earth's limb.
The gamma-ray counts maps of the GMCs are presented in Fig.\ref{fig:counts}.
We generated the source models using make4FGLxml.py\footnote{\url{https://fermi.gsfc.nasa.gov/ssc/data/analysis/user/make4FGLxml.py}}, which consists of the sources in 4FGL-DR2 within the ROI enlarged by 7$^{\circ}$, the galactic diffuse background emission (gll\_iem\_v07.fits), and the isotropic emission background (iso\_P8R3\_SOURCE\_V3\_v1.txt).

\subsection{Spatial analysis and results}
\label{subsec:spatial}
\begin{figure*}[htbp]
    \centering
    \begin{subfigure}[t]{0.3\linewidth}
        \centering
        \includegraphics[width=\linewidth]{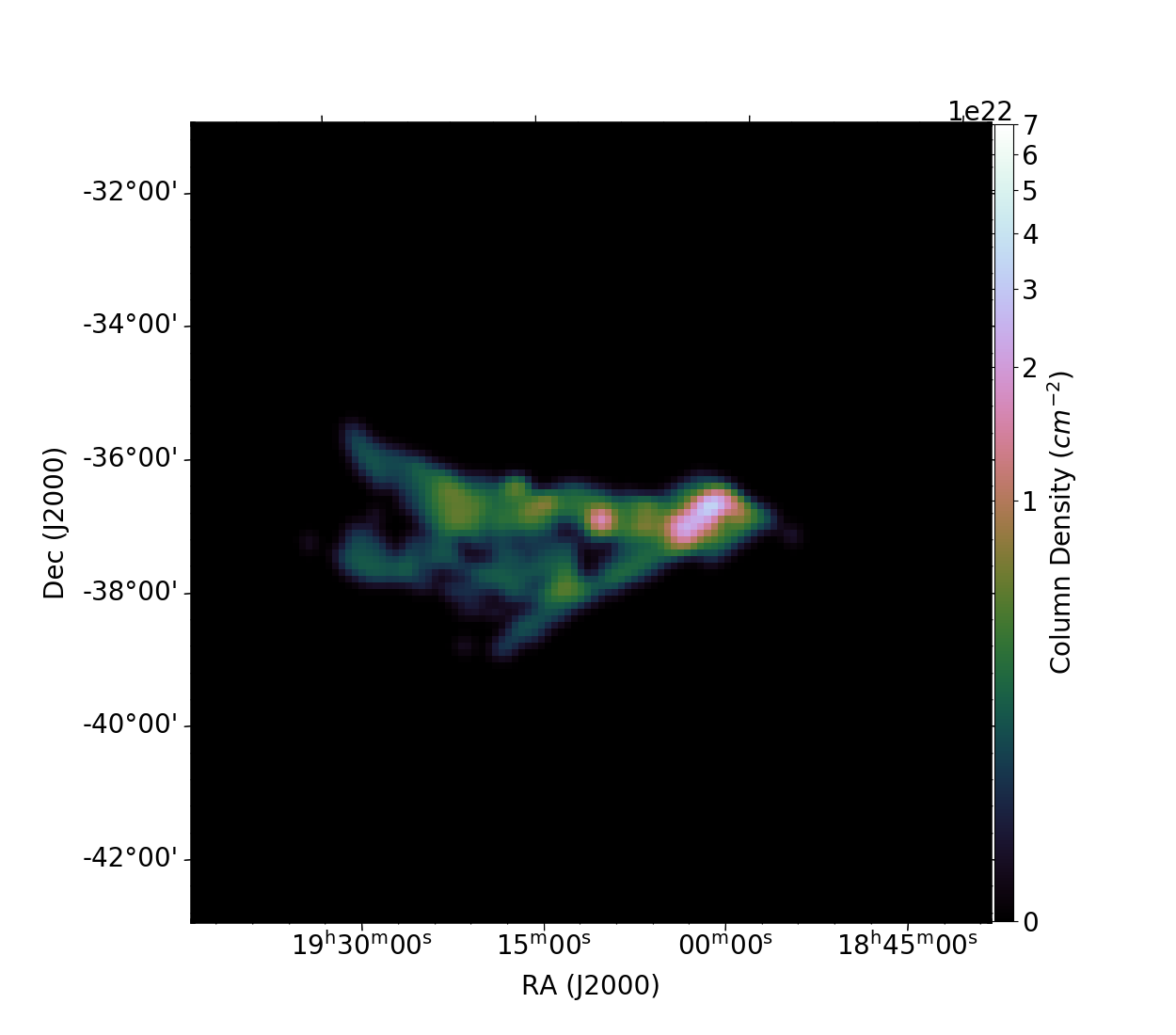}
        \caption{R~CrA}
    \end{subfigure}
    \quad
    \begin{subfigure}[t]{0.3\linewidth}
        \centering
        \includegraphics[width=\linewidth]{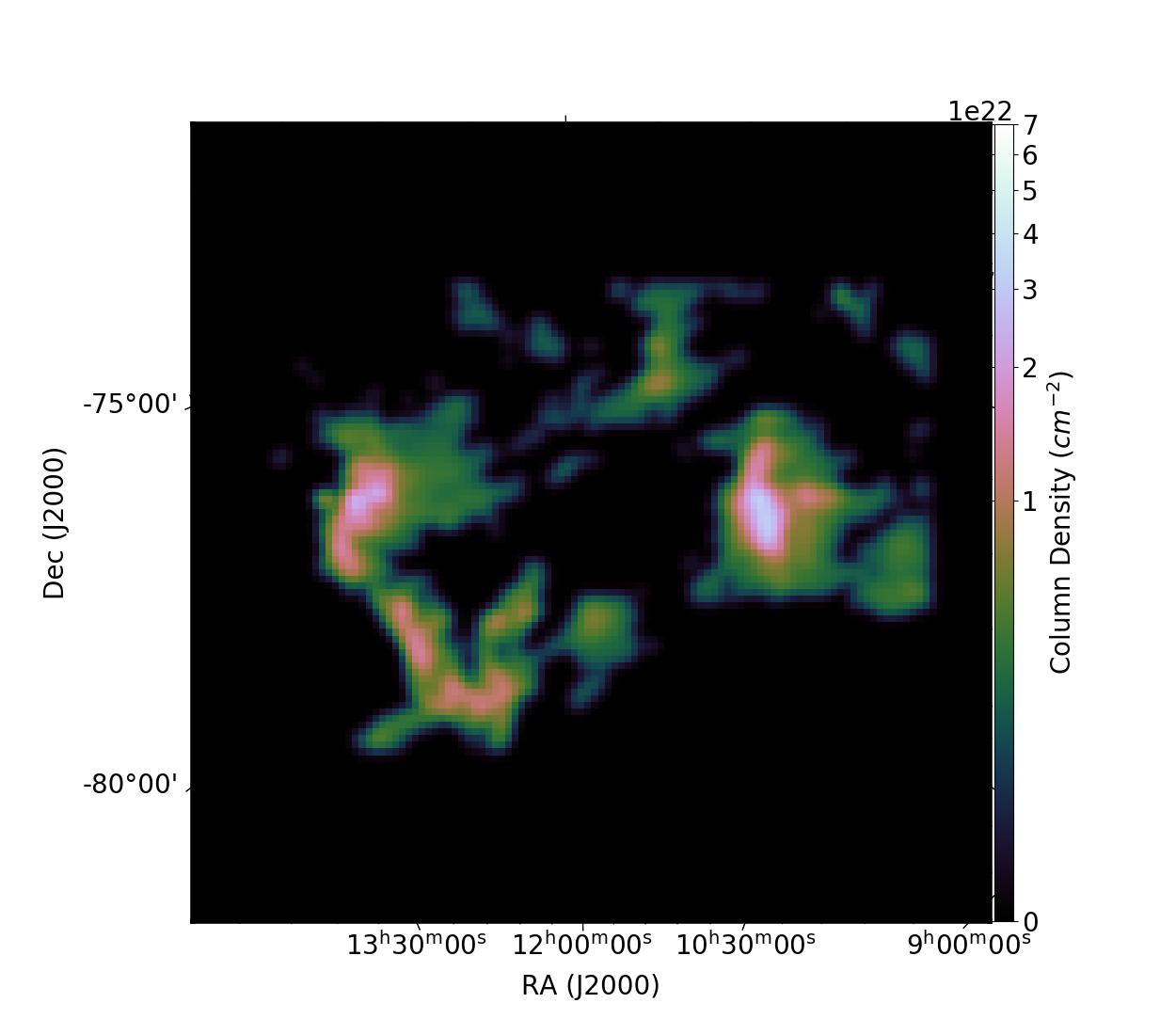}
        \caption{Chamaeleon}
    \end{subfigure}
    \quad
    \begin{subfigure}[t]{0.3\linewidth}
        \centering
        \includegraphics[width=\linewidth]{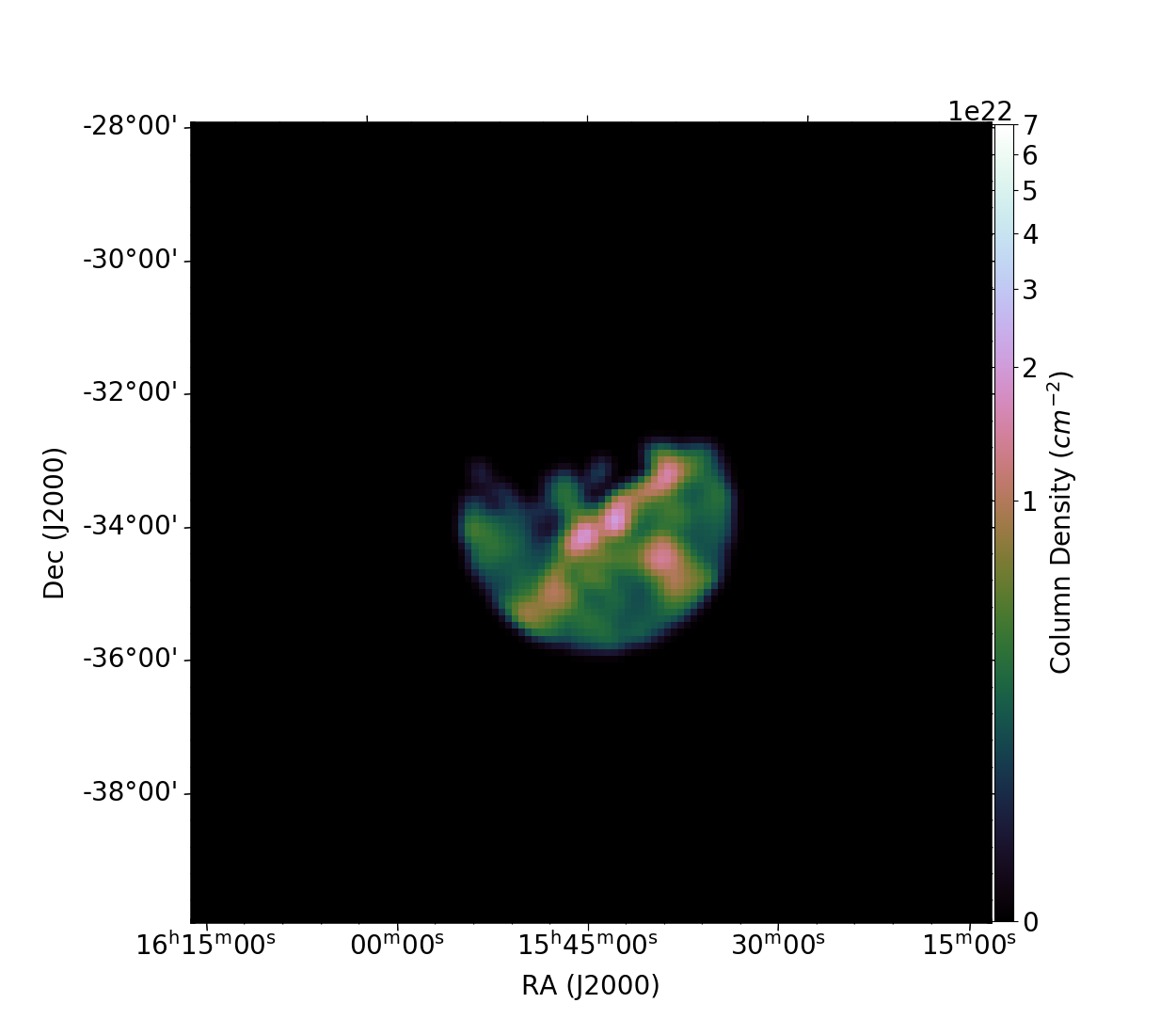}
        \caption{Lupus}
    \end{subfigure}
    \caption{Gas column density maps of the GMCs (in the unit of ${\rm cm}^{-2}$).  Details are described in Sec.\ref{sec:gas}.}
    \label{fig:model}
\end{figure*}

Because the galactic diffuse model mentioned above already included the gamma rays produced in GMCs, so at first,  we generated our own spatial templates to separate the gamma rays associated with the GMCs from the galactic diffuse background.  
Assuming that gamma rays trace the distribution of the molecular gas, we derived the gas column density distribution using data from Planck all-sky observation of thermal dust emission \citep{2013Planck}, and the details are described in Sec.\ref{sec:gas}.
Here, we retained the data with gas column density higher than $10^{21}$~cm$^{-2}$ around the GMCs as the spacial templates of the GMCs (see Fig.\ref{fig:model}).  Meanwhile, the gas column density maps within the ROI enlarged by 10$^{\circ}$ in which the GMCs are subtracted are set as the background templates. 
Besides the gamma rays produced from the pion-decay process induced by the inelastic collision between CR protons and ambient gas,  the diffuse emission  also includes the inverse Compton (IC) scattering of CR electrons  in the interstellar radiation fields (ISRFs).  Thus, We generated  the diffuse IC emission template using GALPROP\footnote{\url{http://galprop.stanford.edu/webrun/}} \citep{galprop} with the parameter set $^SY ^Z4^R30^T150^C2$ in 
 Ref.~\cite{fermi_diffuse_old}.
Moreover, we used the log-parabola function ($dN/dE=N_0 (E/E_{\rm b})^{-\alpha -\beta log(E/E_{\rm b})}$) as the spectral model of the gamma-ray emission associated with the gas templates. 

\begin{figure*}[htbp]
    \centering
    \begin{subfigure}[t]{0.3\linewidth}
        \centering
        \includegraphics[width=\linewidth]{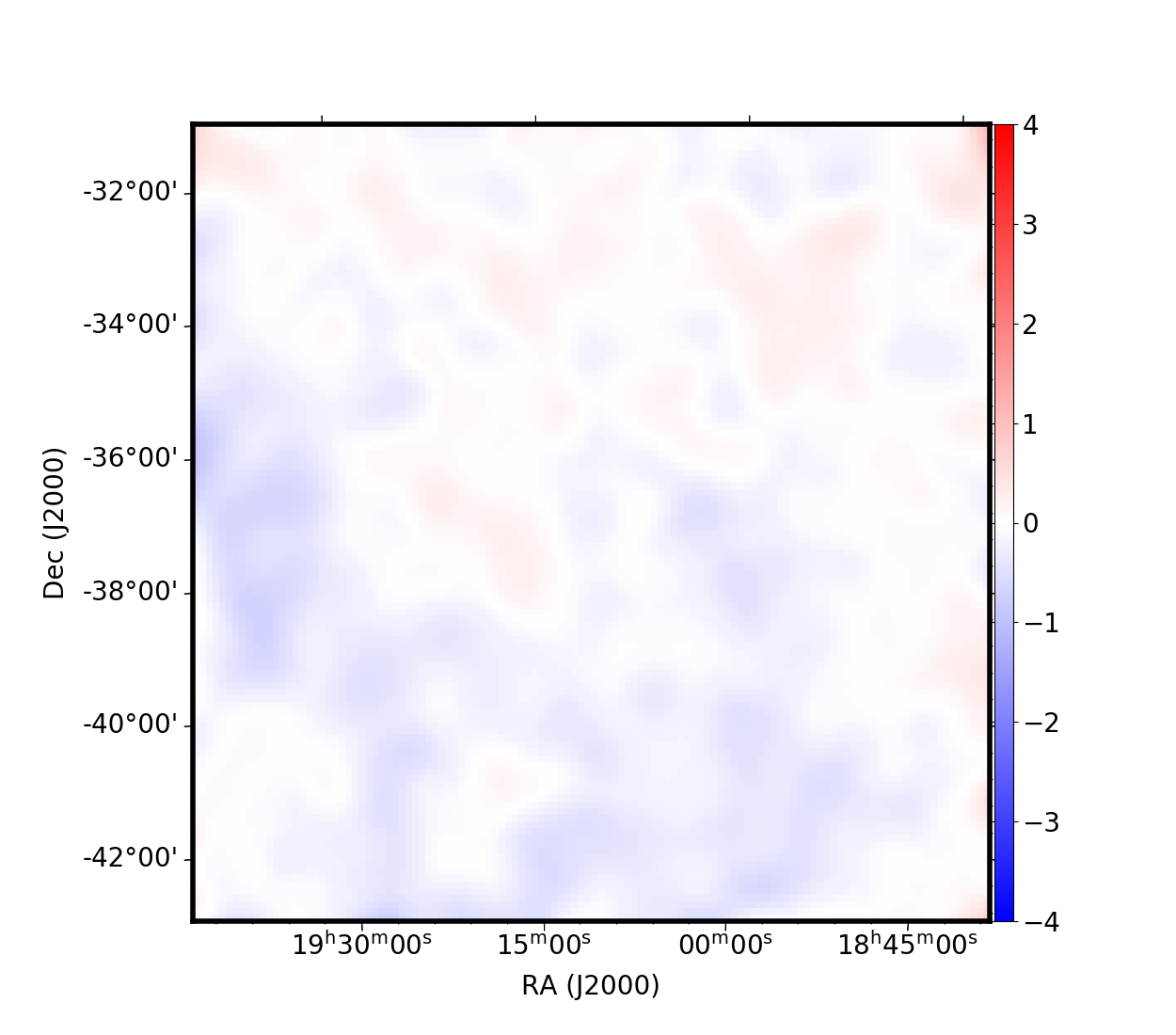}
        \caption{R~CrA}\label{res:1}
    \end{subfigure}
    \quad
    \begin{subfigure}[t]{0.3\linewidth}
        \centering
        \includegraphics[width=\linewidth]{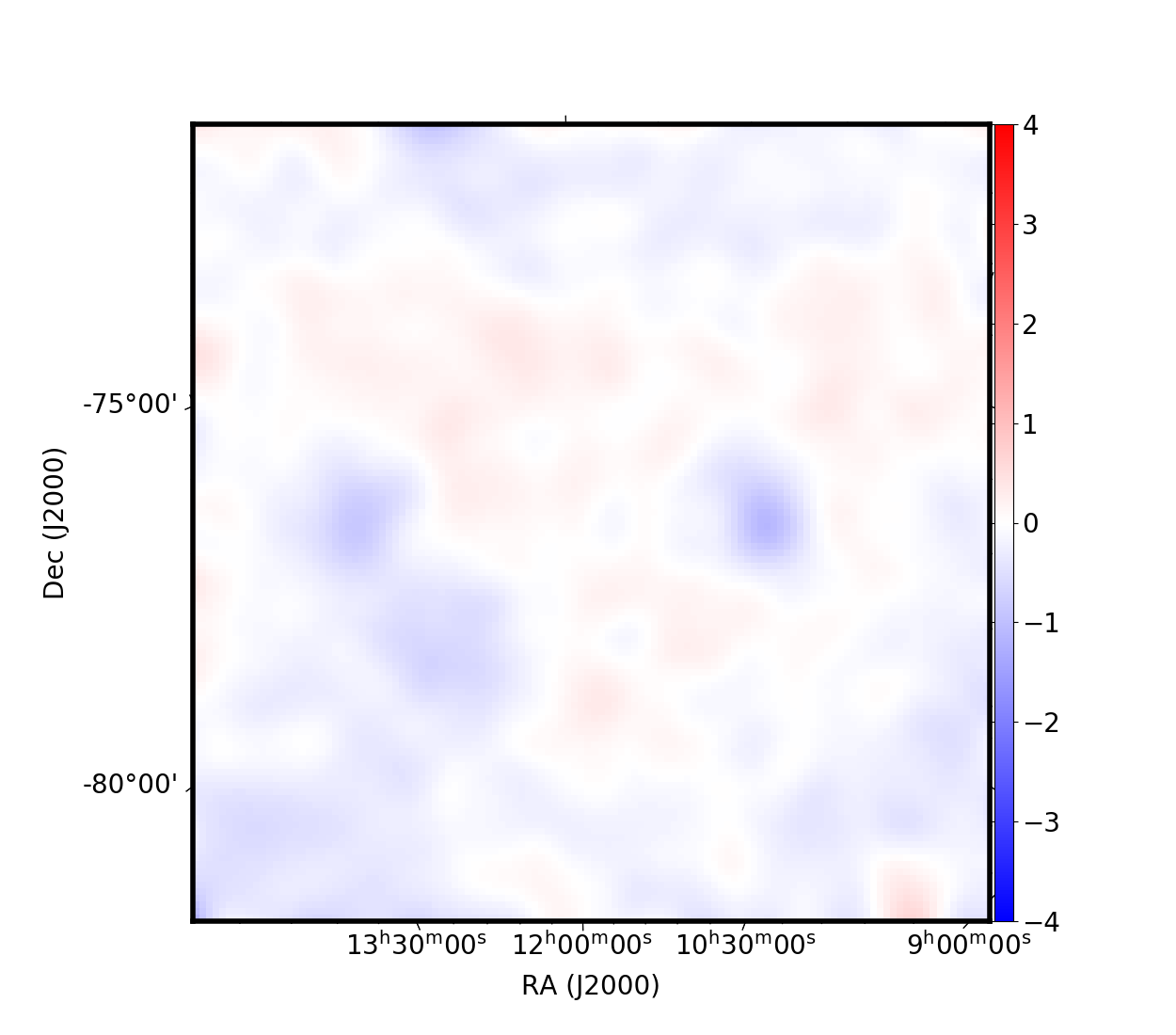}
        \caption{Chamaeleon }\label{res:2}
    \end{subfigure}
    \begin{subfigure}[t]{0.3\linewidth}
        \centering
        \includegraphics[width=\linewidth]{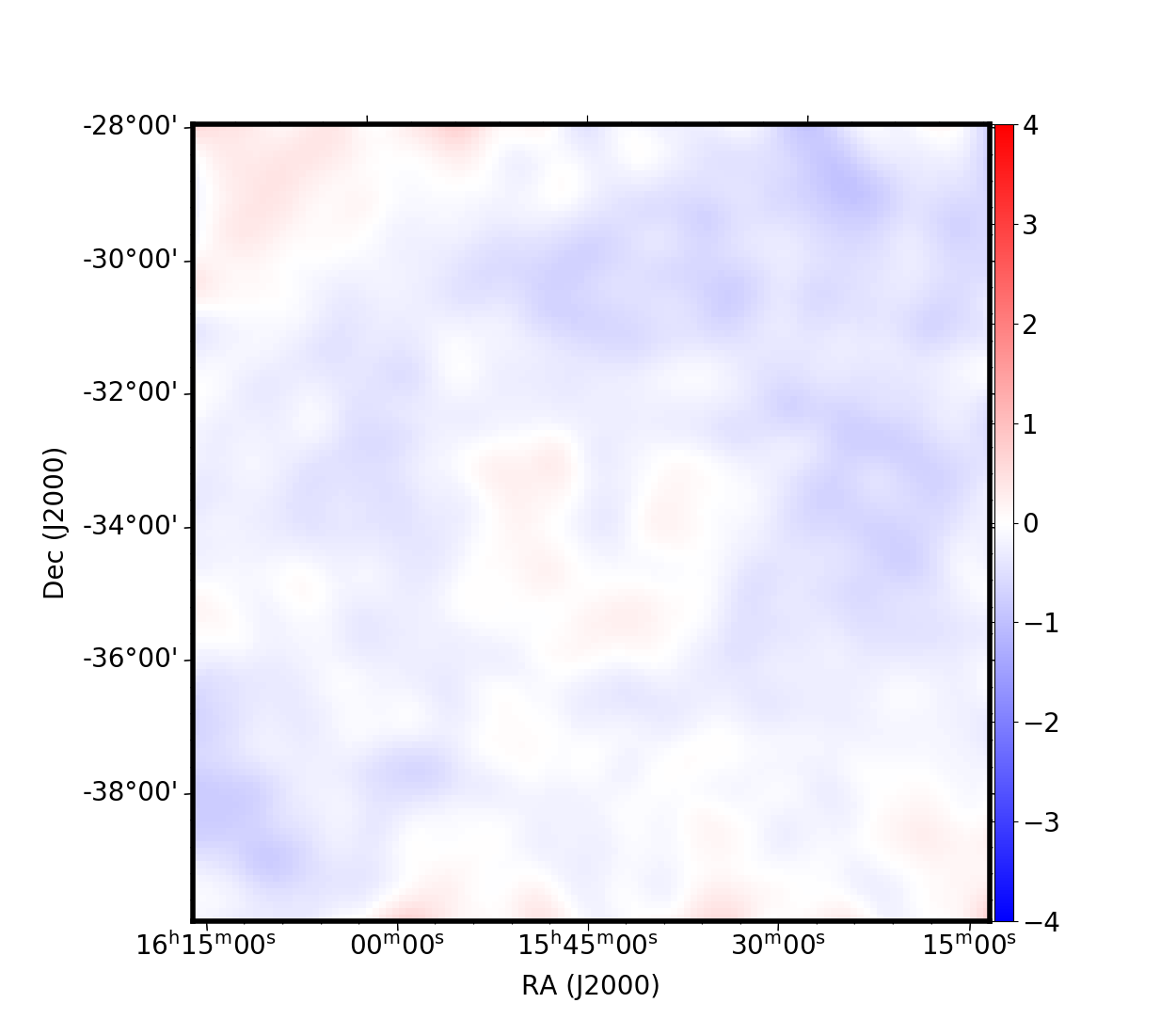}
        \caption{Lupus}\label{res:3}
    \end{subfigure}
    \caption{Residual significance (signal-to-noise, S/N) maps of the GMCs, smoothed with Gaussian filter of 0.3$^{\circ}$.     }
    \label{fig:res}
\end{figure*}

Next, using {\sl gtlike}, we performed likelihood analysis with the modified source models, in which the galactic diffuse background emission is replaced by the emission from the GMCs, the gas background, and the IC component. 
During the fitting process, the normalizations and spectral parameters for diffuse sources mentioned above as well as the sources with significance $\leq 5\,\sigma$ and angular distance $\geq 10^{\circ}$ to the center of each ROI are set free. 
Based on the best-fit results, we generated the model maps using {\sl gtmodel}, and by subtracting the model maps from the observed gamma-ray counts maps we obtained the residuals for the ROI. To check the fitting results statistically, we then divided the residual counts by the square root of the observed gamma-ray counts, i.e., the residual significance ((signal-to-noise, in $\sigma$), of each pixel.
As can be seen from Fig.\ref{fig:res}, for each ROI, the significance of almost all the residual emissions is within $\pm3\sigma$. In other words, our gas templates can approximate the observed gamma-ray distribution of the GMCs well, which allows for further analysis.

\subsection{Spectral analysis and results}

\begin{figure}[htbp]
    \centering
    \includegraphics[width=0.5\linewidth]{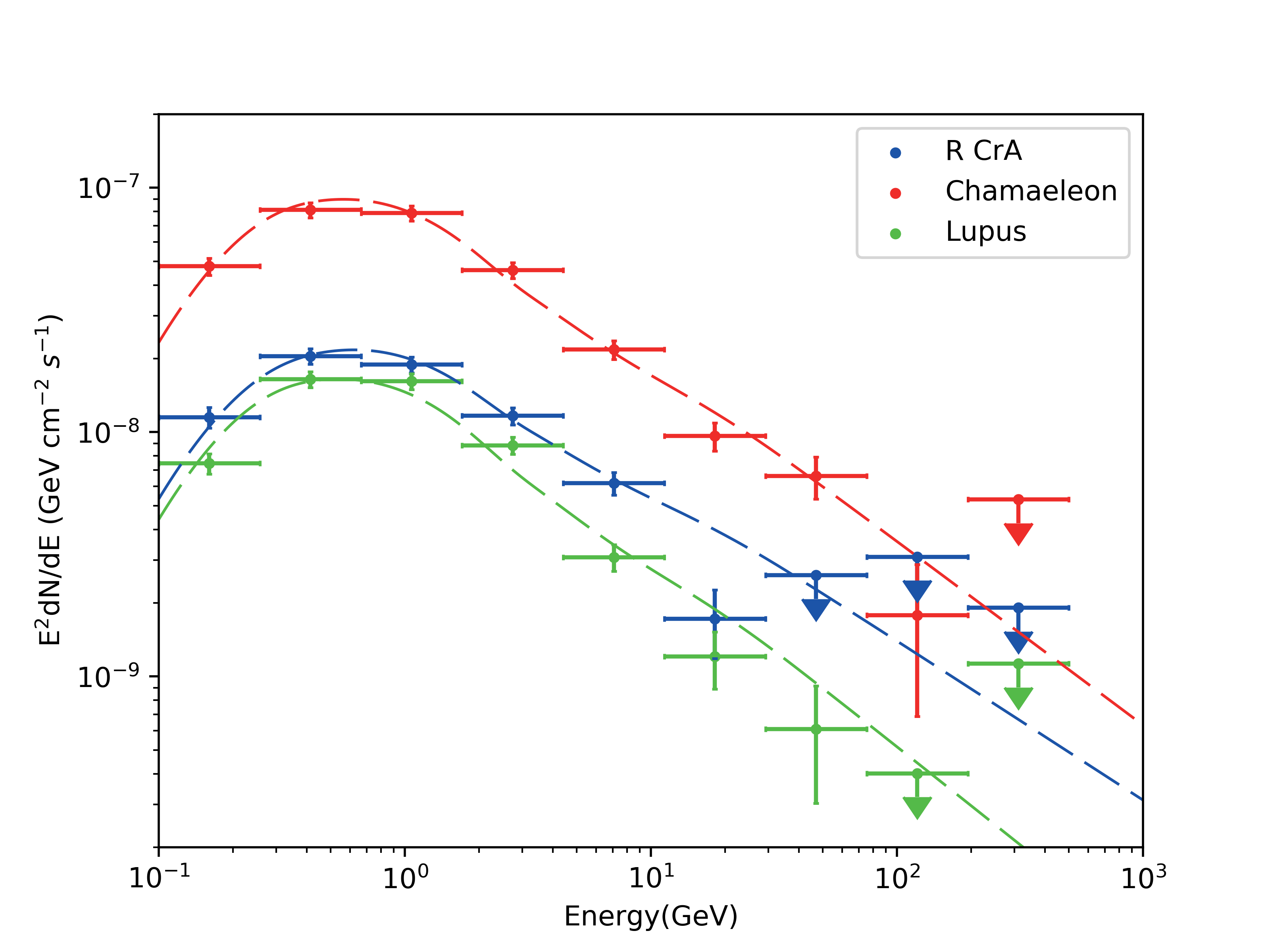}
    \caption{Gamma-ray spectral energy distributions of the GMCs. For each data point, the error bar indicates the uncertainty caused by the statistics and the systematic error. 
    The dashed lines show the corresponding best-fit spectra.}
\label{fig:gamma}
\end{figure}

\section{Gas column density estimation }
\label{sec:gas}

To investigate the property of CRs within the clouds, we also need to estimate the gas column density of the GMCs.
The main components of the clouds, i.e.,  atomic and molecular hydrogen gases are usually traced by  21 cm  \HI\  line and 2.6 mm CO line emission, respectively. 
However, in many cases, only CO and \HI\ observation data are not enough to derive the entire mass of the neutral gas \citep{grenier05}. 
Therefore, we chose the infrared emission from cold interstellar dust to derive the gas column density, which is an independent measurement and will take the untraceable "dark" gas account.

To establish the relation between dust opacity and column density,
we adopted the Eq.(4) from \cite{2014Planck}, in which the dust opacity $\tau_{\rm M}$ is represented as a function of wavelength $\lambda$, 
\begin{equation}\label{equ:dust}
\tau_{\rm M}(\lambda) = \left(\frac{\tau_{\rm D}(\lambda)}{N_{\rm H}}\right)^{\rm dust}[N_{{\rm HI}}+2X_{\rm CO}W_{\rm CO}].
\end{equation}
In the above equation, $(\tau_{\rm D}/N_{\rm H})^{\rm dust}$ is the reference dust emissivity measured in low-$N_{\rm H}$ regions, and $X_{\rm CO}=N_{\rm H_2}/W_{\rm CO}$ is the conversion factor between the integrated brightness temperature of the CO ($W_{\rm CO}$) and the density of molecular hydrogen ($N_{\rm H_2}$). 
According to Eq.\ref{equ:dust},  the total gas density ($N_{\rm H}$) can be estimated as 
\begin{equation}
N_{\rm H} =N_{\rm {HI}}+2 N_{\rm H_2} = \tau_{\rm M}(\lambda)\left[\left(\frac{\tau_{\rm D}(\lambda)}{N_{\rm H}}\right)^{\rm dust}\right]^{-1}. 
\label{equ:2}
\end{equation}
By substituting $(\tau_{\rm D}/N_{\rm H})^{\rm dust}_{353{\rm~GHz}}=(8.4\pm3.0)\times10^{-27}$~cm$^2$ for the dust emissivity at 353~GHz (which is adopted from the Table 4 of \cite{2014Planck}) into Eq.\ref{equ:2},  we derived the gas column density maps for the GMC regions. Then we set the data points as zero while the column density $<10^{21}\,{\rm cm}^{-2}$ to generate spatial templates of the GMCs for {\sl Fermi}-LAT data analysis (see Sec.\ref{subsec:spatial}).  
The gas column density distribution maps of the GMCs are shown in Fig.\ref{fig:model}.

In  previous studies (see, e.g., Ref.~\cite{yang14,yang16}), the reference dust emissivity was considered to be a constant value. However, such a ratio is not uniform in our Galaxy, and the difference can be as high as 100\% as shown in Fig.17 of Ref.~\cite{2013Planck}. 
 To calibrate the ratio between gas column density and dust opacity, we adopted the linear relation between the optical depth and the extinction as in Ref.~\cite{2015Herschel}, which is

\begin{equation}
    A_V=\gamma\tau_{850}+\delta 
    \label{equ:3}
\end{equation}
where $A_V$ stands for the extinction in  $V$-band, 
$\tau_{850}$ stands for the optical depth at frequency $\nu=353$~GHz ($\lambda=850$~$\upmu$m) ,while $\gamma$ and  $\delta$ are constants. Assuming the extinction is proportional to the gas column density, then we can get
\begin{equation}
    N_{\rm H}=N_{\rm HI}+2N_{\rm H_2}=\beta_V{A_ V},
    \label{equ:4}
\end{equation}

in which $\beta_V=5.8\times10^{21}\mathrm{cm}^{-2}\mathrm{mag}^{-1}$ is the mean ratio of total neutral hydrogen to color excess \citep{1978Bohlin}.

\begin{figure*}[htbp]
    \centering
    \begin{subfigure}[t]{0.3\linewidth}
        \centering
        \includegraphics[width=\linewidth]{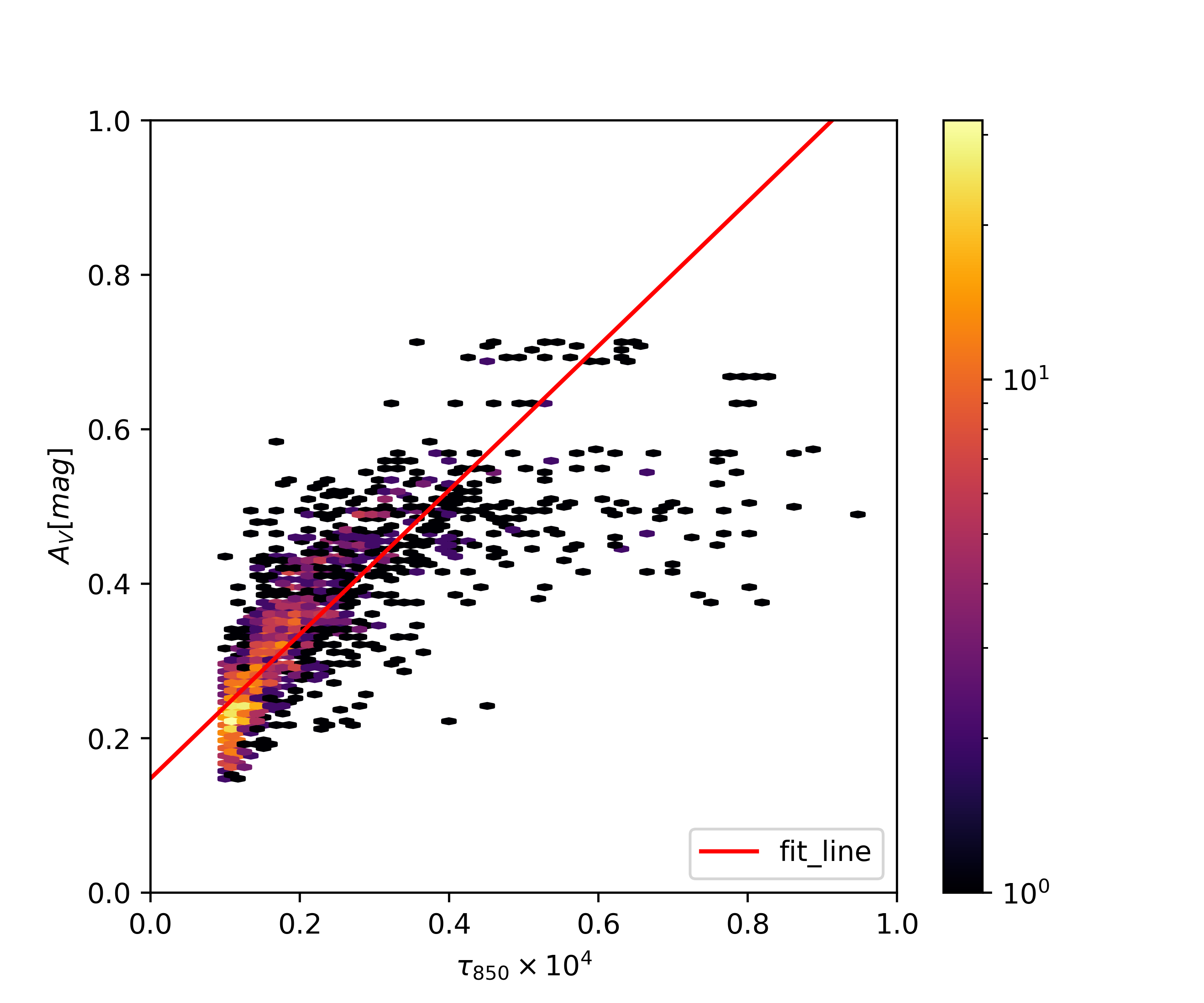}
        \caption{R~CrA}
    \end{subfigure}
    \quad
    \begin{subfigure}[t]{0.3\linewidth}
        \centering
        \includegraphics[width=\linewidth]{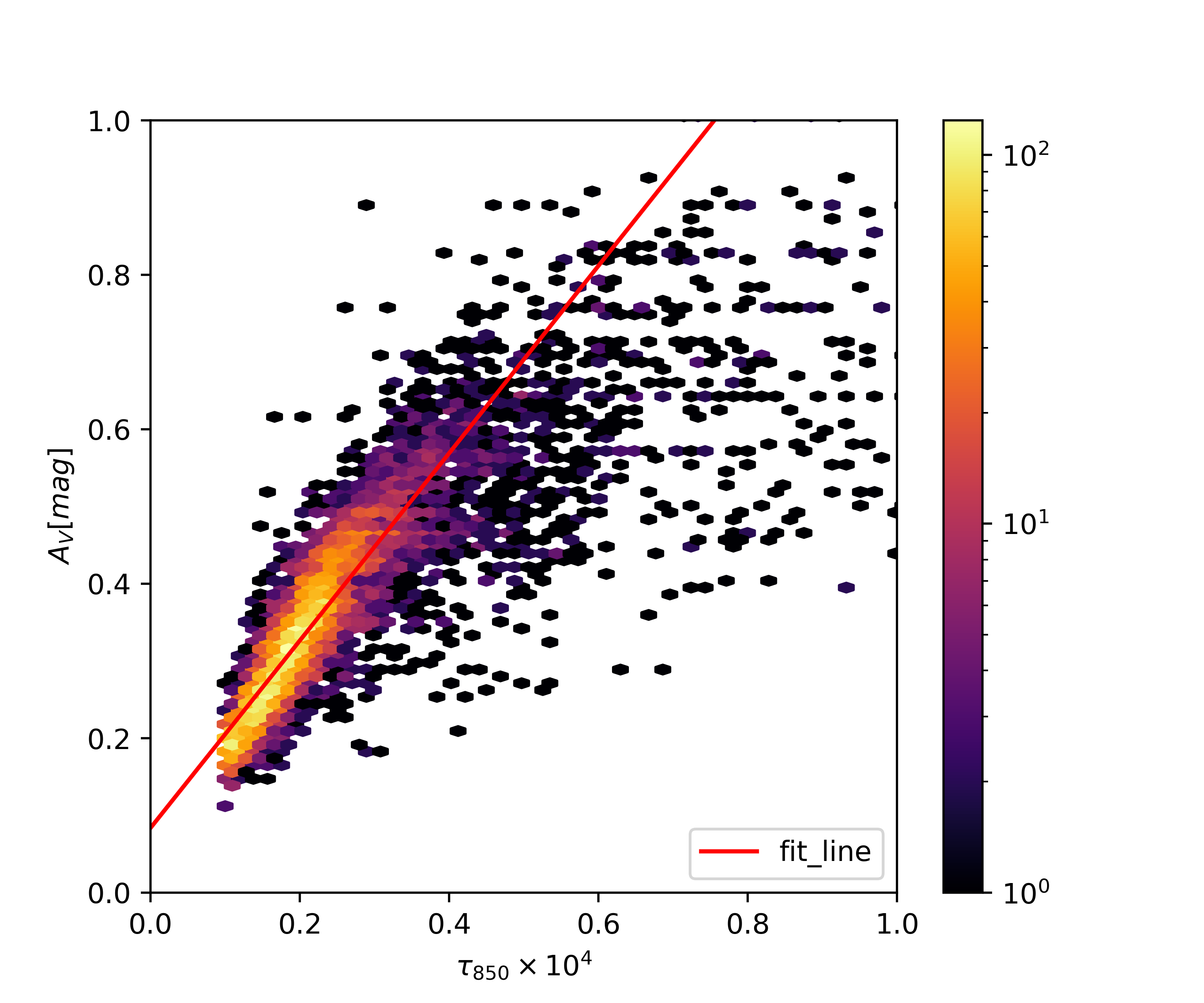}
        \caption{Chamaeleon}
    \end{subfigure}
    \quad
    \begin{subfigure}[t]{0.3\linewidth}
        \centering
        \includegraphics[width=\linewidth]{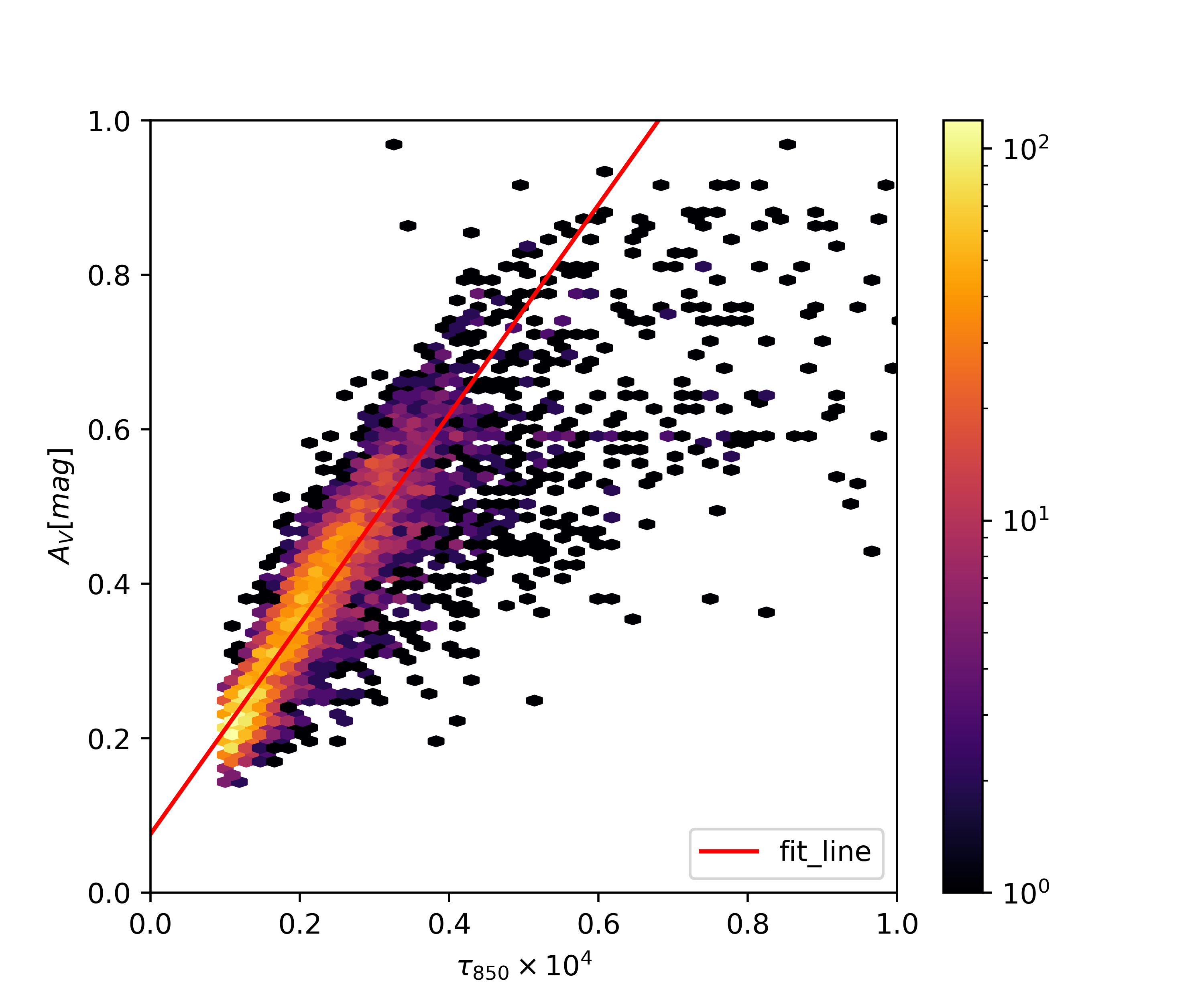}
        \caption{Lupus}
    \end{subfigure}
    \caption{Relationship between the optical depth of given wavelength and the extinction of the GMCs. The color bar shows the counts of data. The red line shows the best-fit result of the linear relation.}
    \label{fig:ext}
\end{figure*}

Combining Eq.\ref{equ:3} and Eq.\ref{equ:4}, now we can use the extinction to calibrate the gas-to-dust ratio for each GMC.
For the extinction, we used the Gaia total galactic extinction map \citep{Gaiamission,GaiaDR3}, where $A_0$ is nearly equivalent to the $V$-band extinction $A_{\rm V}$ within an accuracy of better than 3\% for the extinction range considered in this paper \citep{gaiadr3_cha11}.  
For optical depth, as mentioned in the article \cite{2015Herschel}, we only used the data points whose $\tau_{850}$ was in the range of $0-2\times10^{-4}$.  Both the extinction maps and optical depth maps are generated with the same regions and division of pixels as that of the spatial templates of the GMCs for the gamma-ray data analysis, i.e., Fig.\ref{fig:model}.
Next, we extracted the data points of each pixel from the extinction maps and optical depth maps, and applied maximum likelihood fitting to get the value of $\gamma$ and $\delta$ in Eq.\ref{equ:3} for the GMC regions.  In addition, we calculated $\bar{\tau}$, which is the average $\tau_{850}$ of each GMC region. The best-fit  $\gamma$, $\delta$ as well as $\bar{\tau}$ are listed in Table.\ref{tab:ext}.
As can be seen from Table.\ref{tab:ext}, although $\gamma$ is much larger than $\delta$, the value of $\gamma\bar{\tau}$ and $\delta$ are of a similar order of magnitude. Therefore, $\delta$ can not be ignored.
The best-fit curves representing Eq.\ref{equ:3} as well as the data points of the extinction maps and the dust opacity maps are plotted in Fig.\ref{fig:ext}. 
Using the newly derived $\gamma$ and $\delta$, we recalculated $A_K$, then generated the recalibrated gas column density map by applying Eq.\ref{equ:4} to the same dust opacity data. 

Finally, using the two kinds of gas column density maps, which are derived from a constant gas-to-dust ratio or recalibrated gas-to-dust ratio, we calculated the corresponding total gas masses of the GMCs and listed the results in Table\ref{tab:ext}.

\begin{table*}[htbp]
\caption{Distance and the derived parameters of the GMCs}  
\label{tab:ext}       
\centering                        
\begin{tabular}{cccccc}       
\hline   
GMC& distance & $\gamma$  & $\delta$ & $\left(\tau_{\mathrm{D}} / N_{\mathrm{H}}\right)_{353 \mathrm{GHz}}^{\text {dust }}$&$\left(\tau_{\mathrm{D}} / N_{\mathrm{H}}\right)_{353 \mathrm{GHz}}^{\text {dust }}$\\
& (pc) &($\times10^4$) &($\times10^{-2}$)&($\times10^{-27}$)(C) &($\times10^{-27}$)(R) \\
  \hline
    R~CrA &150& $0.933\pm0.013$ & $14.78\pm0.38$ & $8.4\pm 3.0$ &$12.9\pm0.2$\\
    Chamaeleon&215 & $1.213\pm0.005$ & $8.38\pm0.16$ & $8.4\pm 3.0$&$12.5\pm0.1$\\
    Lupus& 190& $1.358\pm0.008$ & $7.58\pm0.20$ & $8.4\pm 3.0$&$11.3\pm0.1$\\
\hline
\end{tabular}
\\
{\footnotesize The meaning of the fitted parameters, the constant (marked with "C" in table) and recalibrated (marked with "R" in table) gas-to-dust ratio$\left(\tau_{\mathrm{D}} / N_{\mathrm{H}}\right)_{353 \mathrm{GHz}}^{\text {dust }}$ are described in Sec.\ref{sec:gas}.}

\end{table*}

\section{CR properties in the GMCs}
\label{sec:cr}
Assuming the diffuse gamma rays are mainly generated from the pp interaction between CR protons and the gases, then we could derive the spectra of CR in the GMCs, based on their gamma-ray spectra and the derived total gas masses. Here, we adopted the gamma-ray production cross section from Ref.~\cite{2014pp} for calculation, assuming a simple power-law spectrum $F=f_i(E+E_{\rm b})^{\Gamma}$ for the CR protons of each GMC. Here $F$ is the flux density of protons in unit of $\mathrm{cm}^{-2}\, \mathrm{s}^{-1}\,\mathrm{GeV}^{-1}$,  $E$ is the proton energy in unit of GeV, and $\Gamma$ stands for the spectral index. During the fitting process, we fixed $E_{\rm b}$ to 1~GeV since the current \gray data is not very constraining for the low energy part of CR spectra. 
Applying the masses estimated from the two kinds of methods as described in Sec.\ref{sec:gas}, we derived two different normalizations respectively for the CR spectra of the GMCs. For $f_i$(constant), we assumed a constant $\left(\frac{\tau_{\rm D}(\lambda)}{N_{\rm H}}\right)^{\rm dust}$ while for $f_i$(recalibrated) we used the values recalibrated for each GMC, respectively.
The fitted parameters of CR spectra are shown in Table.\ref{tab:cr}.

\begin{table*}[htbp]
\caption{Fitting results of CR spectra for the GMCs}  
\label{tab:cr}       
\centering                        
\begin{tabular}{cccccc}       
\hline    
GMC   & Mass & Mass& $f_i(\times10^{-10})$ & $f_i(\times10^{-10})$  &  $\Gamma$\\
 &(constant)&(recalibrated)&(constant)  & (recalibrated) & \\
 \hline
    R\,CrA &  2231.3&  1450.5&$12.20\pm1.39$ & $18.77\pm2.10$ & $2.716\pm0.051$\\
    Chamaeleon & 14279.5& 9621.2& $19.48\pm1.14$ & $28.89\pm1.69$ & $2.817\pm0.021$\\
    Lupus & 3964.6 & 2957.6& $11.31\pm0.80$ & $15.19\pm1.10$ & $2.868\pm0.049$\\
\hline
\end{tabular}
\\
{\footnotesize The different methods of deriving the mass (in the unit of $M_\odot$)  of each GMC lead to different $f_i$, corresponding to the two panels in Fig.\ref{fig:cr}. }
\end{table*}

\begin{figure*}[htbp]
    \centering
    \begin{subfigure}[t]{0.45\linewidth}
        \centering
        \includegraphics[width=\linewidth]{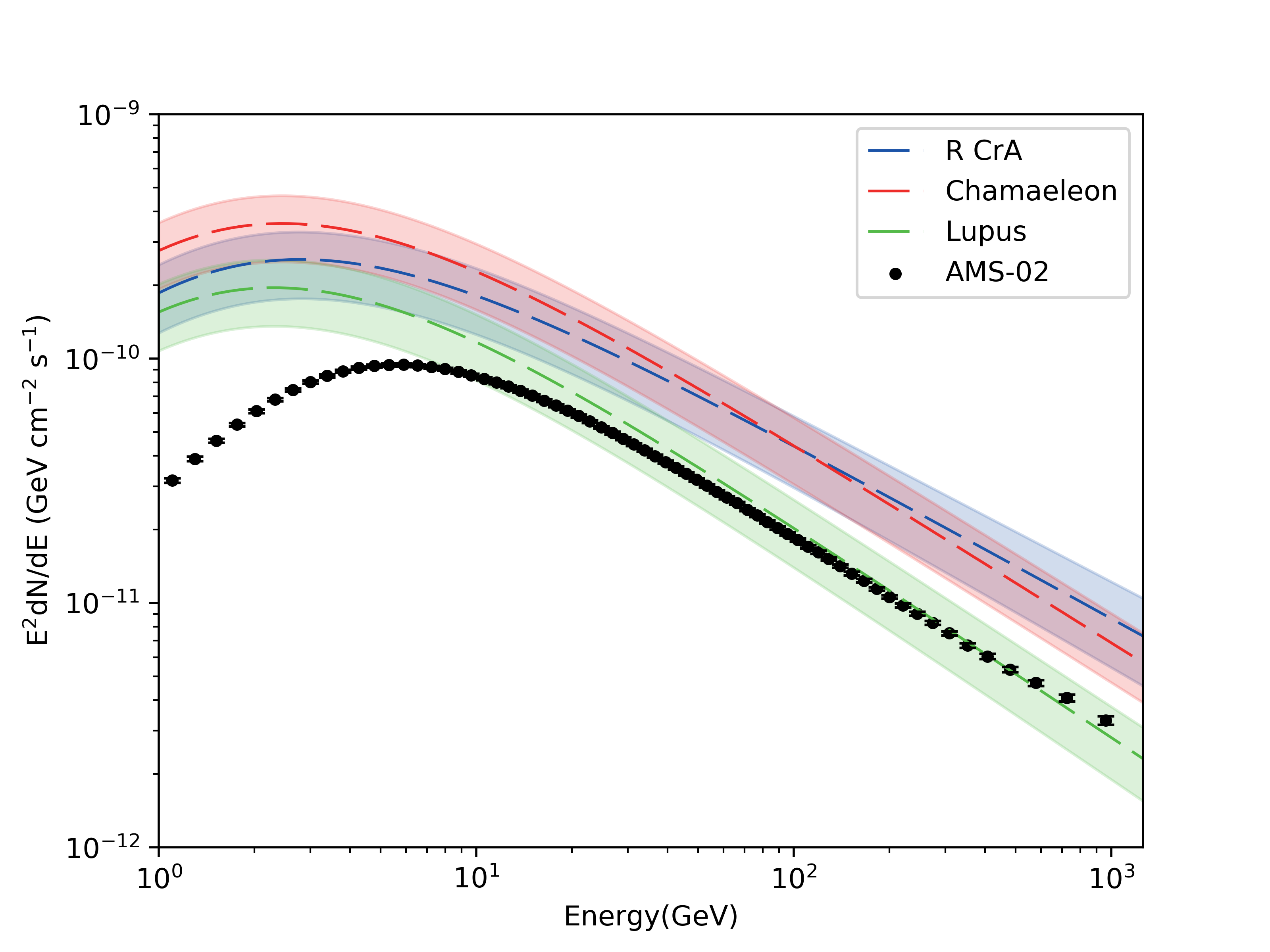}
        \caption{CR spectrum (constant)}\label{cr:a}
    \end{subfigure}
    \quad
    \begin{subfigure}[t]{0.45\linewidth}
        \centering
        \includegraphics[width=\linewidth]{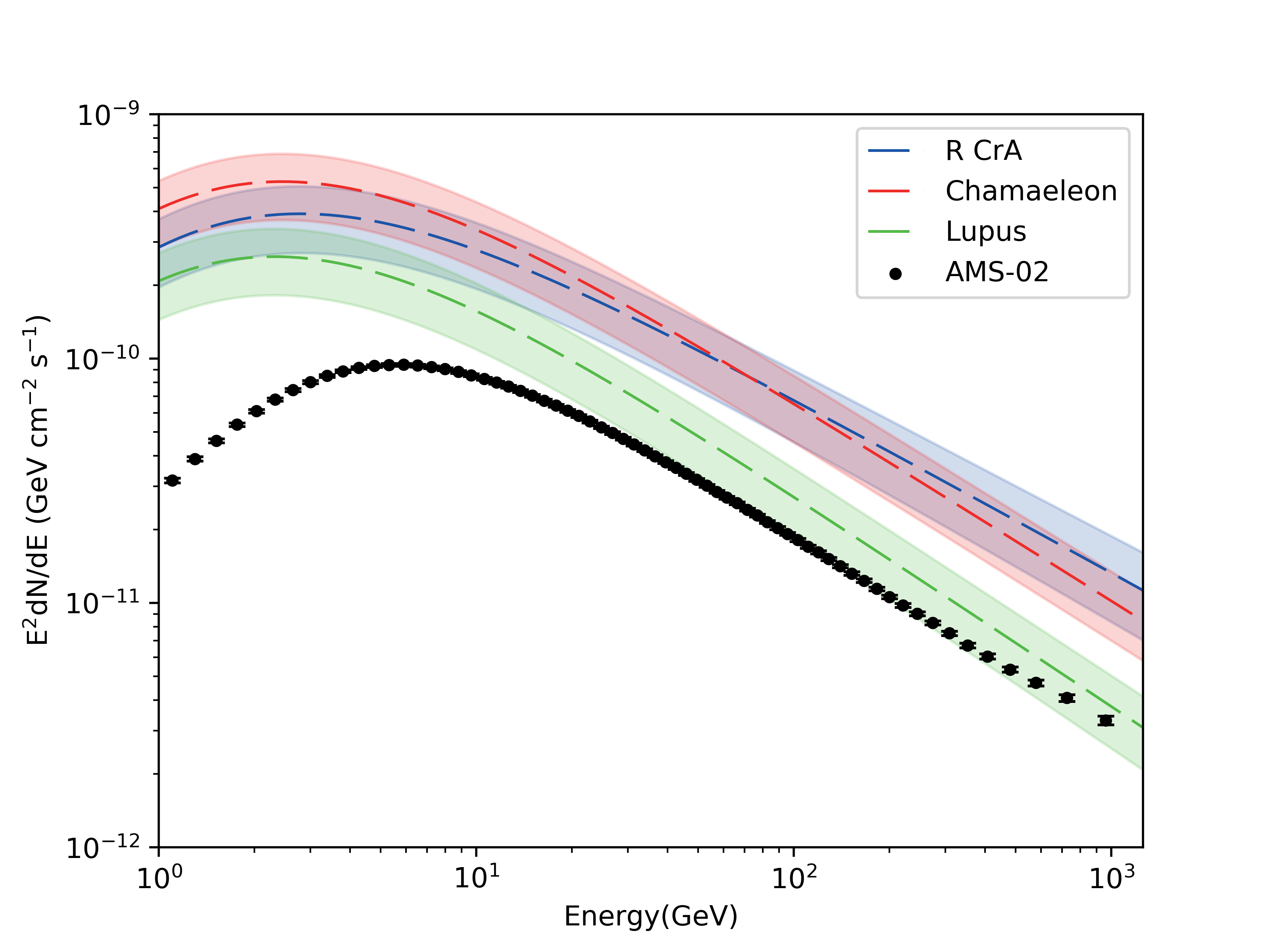}
        \caption{CR spectrum (recalibrated)}\label{cr:b}
    \end{subfigure}
    \caption{Derived CR spectra of the GMCs. The dashed lines show the best-fit CR spectra, the shaded regions represent the 1 $\sigma$ errors including both the statistics and the systematic uncertainty, and the red dots are the AMS-02 observation data acquired from Ref.~\cite{ams02}. The left panel shows the results assuming a constant $\left(\frac{\tau_{\rm D}(\lambda)}{N_{\rm H}}\right)^{\rm dust}$, and the right panel shows the results using the recalibrated values (see Sec.\ref{sec:gas} for details) for each GMC. }
    \label{fig:cr}
\end{figure*}

Next, we plotted the derived CR spectra together with the local measurement by AMS-02 \citep{ams02} in Fig.\ref{fig:cr}. We showed the results with both a constant ratio between the gas column and dust opacity $\left(\frac{\tau_{\rm D}(\lambda)}{N_{\rm H}}\right)^{\rm dust}$ (Fig.\ref{cr:a}) and the ratios recalibrated in Sec.\ref{sec:gas} (Fig.\ref{cr:b}). We found that for both cases the shape of the derived CR spectra are in good consistency with the AMS-02  observation. But the total normalizations show significant differences, even though we have taken into account the $30\%$ uncertainty of the ratio between gas column and dust opacity and the $20\%$ uncertainty in the \gray production cross sections as systematic errors (shown as the shaded area in Fig.\ref{fig:cr}).

\section{Discussion}
\label{sec:dis}
In this paper, we analyzed 12 years of LAT observation data towards three nearby GMCs, and used both the dust opacity map and the extinction map to derive the gas column density map in GMCs. We found the ratio between gas column density and dust opacity can be quite different from cloud to cloud, which may be due to the differences in dust composition and grain distribution \citep{2015Herschel}.
With both the gas distribution and gamma-ray measurement, we derived the CR spectra in all three clouds. We found in all three clouds the spectral shapes are consistent with the measurement of local CRs in the solar system, but the absolute normalizations are significantly different. 

The differences in the ratios between gas column density and dust opacity in different GMCs have strong implications for studying CRs via gamma-ray observations. Since the derived CR density from the gamma-ray flux would be inversely proportional to the derived total mass, which is dependent on such ratios. In the former studies \citep[e.g.,][]{yang14,yang16}, a uniform ratio between gas column density and dust opacity is used in the Galaxy. In this work, we found that the ratio for the nearby GMCs are similar to each other, but can have a difference  of about 50\% compared with the value derived from thew whole sky. Such difference can be even higher when we consider further GMCs, in which not only dust composition and grain distribution but also the dust-to-gas ratio can be quite different. Thus to derive the exact CR distribution in our Galaxy, using only the dust opacity as the gas tracer may bring significant systematic errors in the absolute normalizations of the CR spectra, and more efforts are required to determine the gas distributions.  
We note that the current results rely on the assumption that the dust-to-gas ratios are identical in the three GMCs, which is reasonable considering the proximity of all three clouds. But if such an assumption breaks down, one can not even use the extinction map to calibrate the gas mass. 

The different CR densities derived in this paper may reflect the inhomogeneous distribution of CRs near the solar system. The height above the Galactic plane for the three clouds is comparable, thus the difference cannot be attributed to the CR gradient above the Galactic plane. On the other hand, we note that the CR flux in Chamaeleon, which locates in the anti-Galactic Center (GC) direction, is significantly higher than that in Lupus, which locates towards the GC. Such a difference is similar to the CR anisotropy, whose phase points to the anti-GC direction below 100~TeV\citep{liu19}. Ref.~\cite{liu19} have proposed a nearby CR source in the anti-GC direction may be responsible for such anisotropy. And Ref.~\cite{zhao22} further proposed that the Geminga SNR may be a promising candidate. In this scenario, the higher CR flux in Chameleon can be explained by its vicinity to the local CR source. Further modeling of the CR propagation may be needed to confirm such a scenario.

\acknowledgments
Bing Liu acknowledges the support of the NSFC under grant 12103049. Rui-zhi Yang is supported by  the NSFC under grants 12041305, 11421303 and the national youth thousand talents program in China.

 This work has made use of data from the European Space Agency (ESA) mission
{\it Gaia} (\url{https://www.cosmos.esa.int/gaia}), processed by the {\it Gaia}
Data Processing and Analysis Consortium (DPAC,
\url{https://www.cosmos.esa.int/web/gaia/dpac/consortium}). Funding for the DPAC
has been provided by national institutions, in particular, the institutions
participating in the {\it Gaia} Multilateral Agreement.


\bibliographystyle{JHEP}
\bibliography{biblio.bib}

\end{document}